\documentclass[prl,amsmath,amssymb,nofootinbib,twocolumn,nobibnotes]{revtex4-2}

\usepackage[utf8]{inputenc} 
\usepackage[T1]{fontenc}    
\usepackage{booktabs}       
\usepackage{amsfonts}       
\usepackage{nicefrac}       
\usepackage{microtype}      
\usepackage{graphicx}
\usepackage{natbib}
\usepackage{mathrsfs}
\usepackage{caption}
\usepackage{overpic}
\usepackage{adjustbox}
\usepackage{graphicx}
\usepackage{color}
\usepackage{subcaption} 
\usepackage{float} 
\begin{document}
\title{Probing Dark Matter via the Migdal Effect in LUX-ZEPLIN Data}


\author{Yang Yu}
\affiliation{Key Laboratory of Dark Matter and Space Astronomy, Purple Mountain Observatory, Chinese Academy of Sciences, Nanjing 210023, China}
\affiliation{School of Astronomy and Space Science, University of Science and Technology of China, Hefei 230026, China}

\author{Lei~Feng}
\email{fenglei@pmo.ac.cn (corresponding author)}
\affiliation{Key Laboratory of Dark Matter and Space Astronomy, Purple Mountain Observatory, Chinese Academy of Sciences, Nanjing 210023, China}
\affiliation{School of Astronomy and Space Science, University of Science and Technology of China, Hefei 230026, China}

\begin{abstract}

Sub-GeV dark matter (DM) particles deposit little energy in conventional direct-detection experiments and are therefore difficult to probe, but the Migdal effect provides an electronic signal from atomic ionization accompanying the recoil. 
Using the new results of the LUX-ZEPLIN experiment with a 4.2 tonne-year exposure, we evaluate the sensitivity of the Migdal effect to the spin-independent DM--nucleon scattering cross section. 
For masses between $350~\mathrm{MeV}$ and $10~\mathrm{GeV}$, the projected 90\% upper limits range from $6.5\times10^{-37}$ to $3.1\times10^{-43}~\mathrm{cm^2}$, and become the strongest limits around $1\,\mathrm{GeV}$.
These limits are then used to constrain light scalar- and vector-mediator DM models, reaching a coupling bound of $\alpha\simeq10^{-12}$, stronger than or comparable to those from kaon decays, stellar cooling, and SN1987A.

\end{abstract}
\maketitle

\section{Introduction}\label{sec:intro}

Gravitational evidence for dark matter (DM) is supported by diverse observations, including galactic rotation curves~\cite{Rubin:1980zd}, galaxy-cluster dynamics~\cite{Zwicky1937}, the cosmic microwave background~\cite{Planck2018}, and large-scale structure~\cite{Clowe:2006eq}. 
Nevertheless, the non-gravitational interactions of DM and its mass scale remain undetermined. 
Direct detection experiments typically search for nuclear recoils produced by halo DM scattering off target nuclei~\cite{GoodmanWitten1985, UndagoitiaRauch2016, BertoneHooperSilk2005, Billard2022}. 
Current liquid-xenon experiments such as LZ~\cite{LZ2023,LZ2025}, XENON~\cite{XENONnT2023} and PandaX~\cite{PandaX4T2021} have made substantial progress, with their spin-independent sensitivities approaching the neutrino floor~\cite{OHare2021}.

However, for DM masses well below the target-nucleus mass, the maximum nuclear-recoil energy decreases rapidly with DM mass, so that conventional nuclear-recoil searches are severely limited by energy thresholds.
To overcome this limitation, some mechanisms that boost DM to higher velocities, such as cosmic-ray-boosted DM~\cite{BringmannPospelov2019,Yu:2025bwu} and solar-reflected DM~\cite{AnPospelovPradlerRitz2018}, have been proposed.
Beyond direct searches, astrophysical observations provide complementary sensitivity to this mass range.
Indirect detection can probe this regime through searches for the products of DM annihilation or decay, primarily gamma rays on galactic scales~\cite{Wang:2023sxr,Wang:2026khh} and, on cosmological scales, the cosmic microwave background~\cite{Madhavacheril2014} and the 21-cm signal~\cite{Damico2018}.

The Migdal effect is the ionization or excitation of atomic electrons induced by the sudden recoil of a nucleus during a scattering event.
It was introduced for $\alpha$- and $\beta$-decay~\cite{Migdal1941} and later applied to DM scattering~\cite{Ibe2018,Bell2022,Xu:2023wev}.
The emitted electron and the accompanying X-rays and Auger electrons can deposit visible energy beyond the nuclear recoil itself.
For sub-GeV DM, its advantage is that the maximum nuclear recoil is small, while the Migdal electron can still produce a detectable electronic signal.
Experiments such as XENON1T~\cite{XENON1T2019}, DarkSide-50~\cite{DarkSide:2022dhx}, PandaX-4T~\cite{Qiao:2023pbw} and JUNO~\cite{Scopel:2026aqk} have searched for this effect and used it to constrain sub-GeV DM. 
It has now been directly observed in neutron--nucleus collisions by a Chinese collaboration at $5\sigma$~\cite{Yi2026}.
Since the Migdal signal can be visible below the conventional nuclear-recoil threshold, it extends the sensitivity of liquid-xenon experiments to lower DM masses \cite{Dolan2018}.

In this work we use the LZ Migdal signal templates and the corresponding keV-scale analysis framework to obtain spin-independent cross-section upper limits at different DM masses through a Poisson analysis.
On this basis we further consider sub-GeV DM models with light mediator, retaining the full momentum dependence of the propagator, and map the cross-section limits into the coupling combination $\alpha_{med}=g_Ng_\chi/(4\pi)$ as a function of $m_{med}$ for both scalar and vector mediators, comparing with external constraints from astrophysics and particle physics.

This paper is organized as follows: Sec.~\ref{sec:model} introduces the DM model and  the scattering of DM particles in the detector; Sec.~\ref{sec:migdal} presents the Migdal effect and the resulting Migdal cross-section limits from LZ; Sec.~\ref{sec:mediator} shows the coupling limits for scalar and vector mediators; and Sec.~\ref{sec:conclusion} summarizes our conclusions.

\section{DM model}\label{sec:model}

We consider two minimal models in which DM couples to nucleons via a mediator that is sufficiently light and therefore cannot be integrated out. 
For a real gauge-singlet scalar \(S\) or a \(U(1)\) gauge boson \(V_\mu\), the interactions with fermionic DM \(\chi\) and with nucleons \(N\) are given by
\begin{align}
  \mathcal{L}_S &\supset -\,g_\chi S\,\bar{\chi}\chi - g_N S\,\bar{N}N , \label{eq:lag-scalar}\\
  \mathcal{L}_V &\supset -\,g_\chi V_\mu\,\bar{\chi}\gamma^\mu\chi - g_N V_\mu\,\bar{N}\gamma^\mu N , \label{eq:lag-vector}
\end{align}
where \(g_\chi\) denotes the coupling to DM and \(g_N\) the effective coupling to nucleons.
Because the two Lagrangians have the same structure, the leading non-relativistic DM--nucleon interaction is in both cases momentum-independent and spin-independent (SI). 
The coupling is coherent over the nucleus, so that the nuclear cross section carries the usual $A^2$ enhancement.

Within the spin-independent (SI) scattering framework, the DM-nucleon scattering strength is conveniently characterized by the nucleon-level cross section $\sigma_{\mathrm{SI}}$,
\begin{align}
  \sigma_{\mathrm{SI}} &= \frac{(g_N g_\chi)^2\,\mu_{\chi p}^2}{\pi\,(m_{\mathrm{med}}^2+q^2)^2}.
  \label{eq:sigma-SI}
\end{align}
where $m_\chi$ and $m_p$ are the DM and nucleon masses, respectively, and $\mu_{\chi p}=m_\chi m_p/(m_\chi+m_p)$ is the DM-nucleon reduced mass. 
In the non-relativistic limit, the maximum nuclear-recoil energy for an incident DM particle of speed $v$ is $E_R^{\rm max}={2\mu_{\chi N}^2v^2}/{m_N}$, where $\mu_{\chi N}=m_{\chi} m_N/(m_{\chi}+m_N)$.
For $m_{\chi}\ll m_N$, $\mu_{\chi N}\simeq m_{\chi}$ and $E_R^{\rm max}$ decreases as $m_{\chi}^2$. This is the primary reason why low-mass DM is difficult to detect via the conventional nuclear-recoil channel.

The differential nuclear-recoil event rate is given by the convolution of the velocity distribution with the differential cross section~\cite{LewinSmith}:
\begin{equation}
  \frac{dR_{\rm NR}}{dE_R}=
  \frac{\rho_\chi}{m_{\chi} m_N}
  \int_{v_{\min}}^{v_{\max}} d^3v\,
  v\, f(\mathbf v)\,
  \frac{d\sigma}{dE_R},
  \label{eq:rate}
\end{equation}
where $\rho_\chi$ is the local DM density, $f(\mathbf v)$ is the DM velocity distribution in the detector frame, and $v_{\min}=\sqrt{m_N E_R/(2\mu_{\chi N}^2)}$ is the minimum speed required for a nuclear recoil of energy $E_R$.
 
We adopt the standard halo model (SHM), in which the DM velocity distribution in the detector frame follows a truncated Maxwellian form with a characteristic velocity $v_0=220\,\mathrm{km/s}$, an escape speed $v_{\rm esc}=544\,\mathrm{km/s}$, and a mean Earth velocity of about $232\,\mathrm{km/s}$~\cite{LewinSmith}. 
The elastic differential cross section is given by
\begin{equation}
  \frac{d\sigma}{dE_R}=
  \frac{\sigma_N}{E_R^{\rm max}}\,F(q)^2,
  \qquad
  \sigma_N=\sigma_{SI}\left(\frac{\mu_{\chi N}}{\mu_{\chi p}}\right)^2 A^2,
\end{equation}
where $A$ is the mass number of the target nucleus and $F(q)$ is the nuclear form factor. For SI interactions, the Helm form factor is commonly adopted~\cite{Helm1956},
\begin{equation}
  F(q)=\frac{3j_1(q r_n)}{q r_n}\exp\left[-\frac{(q s)^2}{2}\right],
\end{equation}
where $j_1$ is the spherical Bessel function, $s\simeq0.9\,\mathrm{fm}$, and $r_n\simeq1.14A^{1/3}\,\mathrm{fm}$.

\section{Migdal effect in LZ}\label{sec:migdal}
\subsection{Migdal electron signal}

For a given nuclear-recoil energy $E_R$, the visible energy of the Migdal signal is determined by the electron energy, the shell binding energies, and the subsequent atomic de-excitation processes. The event rate can be written conceptually as \cite{Ibe2018}
\begin{equation}
  \frac{dR}{dE_e}=\int dE_R\,
  \frac{dR_{\rm NR}}{dE_R}\,
  P_{\rm Migdal}(E_e\mid E_R),
  \label{eq:migdal-rate}
\end{equation}
where $dR_{\rm NR}/dE_R$ is the nuclear-recoil spectrum and $P_{\rm Migdal}$ is the conditional probability of producing electron energy $E_e$ for a given recoil. 

In the dipole approximation, the differential ionization probability of the $nl$ energy level for a nuclear speed $v_N$ factorizes as~\cite{Ibe2018,Dolan2018}
\begin{equation}
  \frac{dP_{nl}}{dE_e}\simeq
  \frac{q_e^2}{2\pi}\,p_{nl}(E_e),
  \qquad
  q_e\simeq m_e v_N,
  \label{eq:dipole}
\end{equation}
where $m_e$ is the electron mass and $p_{nl}(E_e)$ is an atomic-structure kernel that is independent of the nuclear speed in this approximation. This scaling relation allows the Migdal ionization probability to be estimated directly from the nuclear-recoil energy and underlies the construction of signal templates for low-mass DM searches. 
The signal templates used in this work are constructed using the xenon single-ionization probability tables of Ibe \textit{et al.}~\cite{Ibe2018}, for which the single-ionization probabilities were calculated for all occupied energy levels from $1s$ to $5p$. 
The improved atomic calculations of Cox \textit{et al.}~\cite{Cox2023} can serve as a refined theoretical input.

\begin{figure}[htbp]
  \centering
  \includegraphics[width=0.45\textwidth]{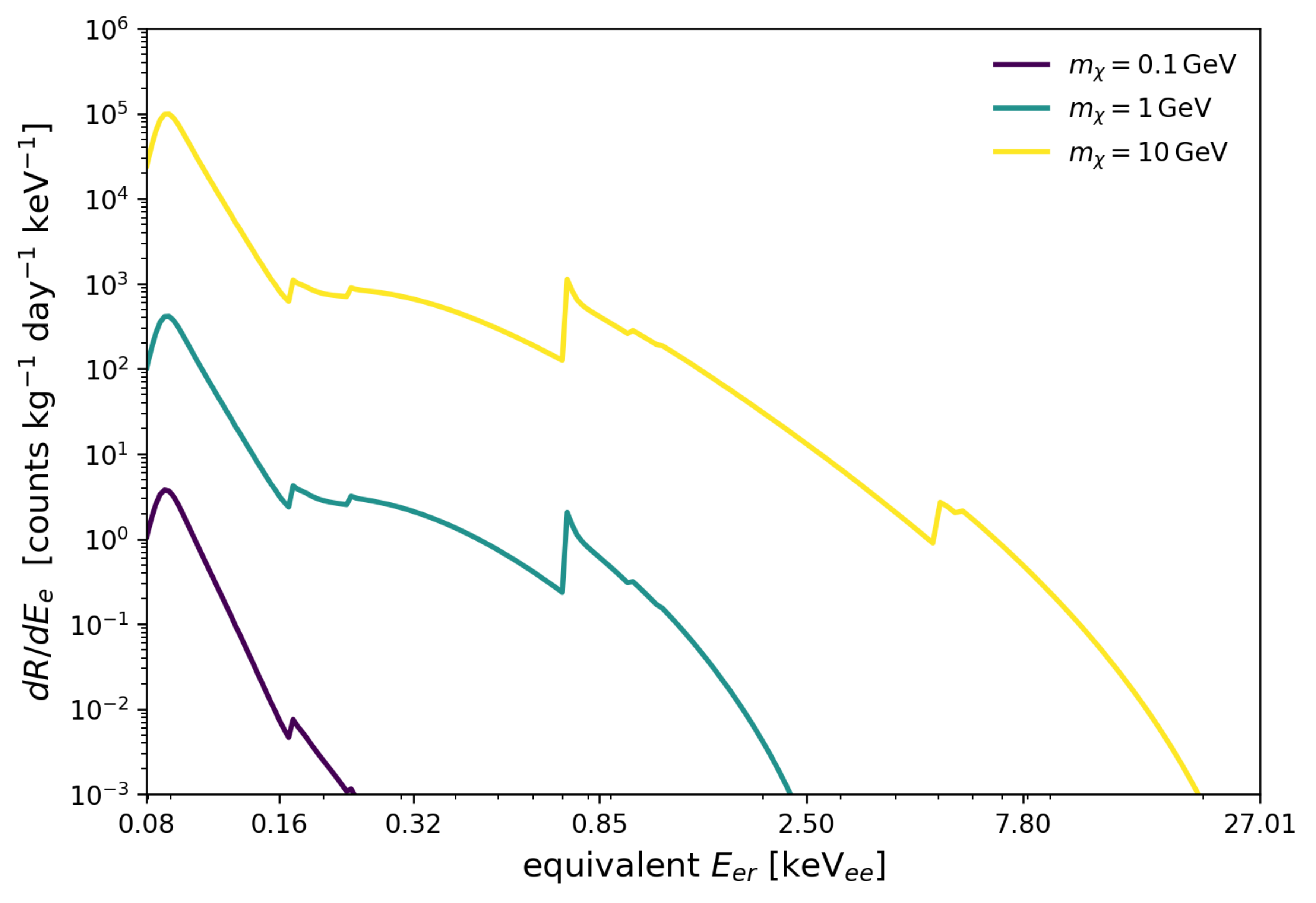}
  \caption{Xe Migdal event rate with cross section $\sigma_n=10^{-35}\,\mathrm{cm^2}$ and $m_{\chi}=0.1,~1,~10\,\mathrm{GeV}$. }
  \label{fig:migdal-rate}
\end{figure}

To compute the rate, we sum over all occupied energy levels and express
Eq.~(\ref{eq:migdal-rate}) as an explicit double integral:
\begin{equation}
\begin{split}
&\frac{dR}{dE_e} = \frac{\rho_\chi}{m_\chi m_N}
  \sum_{nl}\int_0^{E_R^{\max}(m_\chi,v_{\max})} dE_R \times \\
  &\quad \int_{v_{\min}^{\rm ME}(E_e,E_R)}^{v_{\max}} dv\,
  v\,f(\mathbf v)\,\frac{d\sigma}{dE_R}\,
  \frac{q_e^2}{2\pi}\,p_{nl}(E_e-E_{nl}^{\rm b})
\end{split}
\end{equation}
where the minimum DM speed required for a Migdal transition is
\begin{equation}\label{eq:vmin-ME}
  v_{\min}^{\rm ME}(E_e,E_R)=
  \frac{m_N E_R+\mu_{\chi N}E_e}
       {\mu_{\chi N}\sqrt{2m_N E_R}},
\end{equation}
where the dipole momentum transfer is $q_e=m_ev_N=m_e\sqrt{2E_R/m_N}$, and $E_{nl}^{\rm b}$ is the binding energy of the $nl$ energy level. 
For a given deposited energy $E$, the ionization energy available for the $nl$ level is $E_e=E-E_{nl}^{\mathrm{b}}$, and energy levels with $E_e<0$ have no contribution.

The resulting differential event rate, $dR/dE_{e}$, is shown in
Fig.~\ref{fig:migdal-rate}. The differential event rate is concentrated at the lowest energies and falls steeply, while both its normalization and kinematic endpoint increase rapidly with the DM mass. 
The step-like features at $\sim 0.3$ and $2.5~\mathrm{keV}_{ee}$ arise from the successive opening of deeper
xenon shells in the Migdal ionization probability.

\subsection{Method and results}\label{sec:method}

The signal templates are generated for 20 log-uniform DM masses from $100\,\mathrm{MeV}$ to $10\,\mathrm{GeV}$ and normalized to $\sigma_{0}=10^{-37}\,\mathrm{cm^2}$, so that the expected rate scales linearly with $r_S=\sigma_{SI}/\sigma_{0}$. 
The ionization probability is obtained by convolving the dipole factor of Eq.~(\ref{eq:dipole}) with the atomic tables of~\cite{Ibe2018} for all occupied xenon energy levels ($1s$--$5p$), and is converted to a count spectrum in $\mathrm{keV_{ee}}$ using the liquid-xenon electronic-recoil charge yield and detection efficiency. 
In the analysis window $1.0$--$4.0\,\mathrm{keV_{ee}}$, the signal is dominated by the outer $3d$--$5p$ shells, since the binding energies of the inner $1s,2s,2p$ shells exceed the available electron energy.

With an exposure of $4.2$ tonne-year~\cite{LZ2025} and Asimov data $n_i=\mathcal{E}b_i$, we jointly fit the background normalization $r_B$, the signal strength $r_S$, and a signal-normalization factor $f_0$, the latter carrying a Gaussian prior with mean $1$ and width $0.2$. 
The per-bin background rates are assigned a 30\% relative uncertainty and integrated out with 16th-order Gauss--Hermite quadrature.
Conditional on the background rates $\tilde b_i$, the Poisson likelihood reads
\begin{equation}
  \mathcal{L}(\theta)=\prod_i
  \frac{\lambda_i(\theta)^{n_i}}{n_i!}
  \exp[-\lambda_i(\theta)], 
  \label{eq:likelihood}
\end{equation}
where $\lambda_i=\mathcal{E}\left(r_B\tilde b_i+r_Sf_0s_i\right)$, $s_i$ is the signal rate per unit $r_S$ and $\theta=(r_B,r_S,f_0)$ collects the fitted parameters. 
We sample $\theta$ using a Metropolis--Hastings chain, discarding the first 4000 burn-in steps and retaining $20\,000$ unthinned samples. 
For each DM mass, we report $\sigma_{\mathrm{SI}}^{90}=q_{0.9}(r_S)\times10^{-37}\,\mathrm{cm^2}$, where $q_{0.9}(r_S)$ denotes the 90th percentile of the marginal posterior distribution of $r_S$. 
The resulting limit is therefore a Bayesian posterior quantile rather than a frequentist profile-likelihood interval.

\begin{figure}[htbp]
  \centering
  \includegraphics[width=0.45\textwidth]{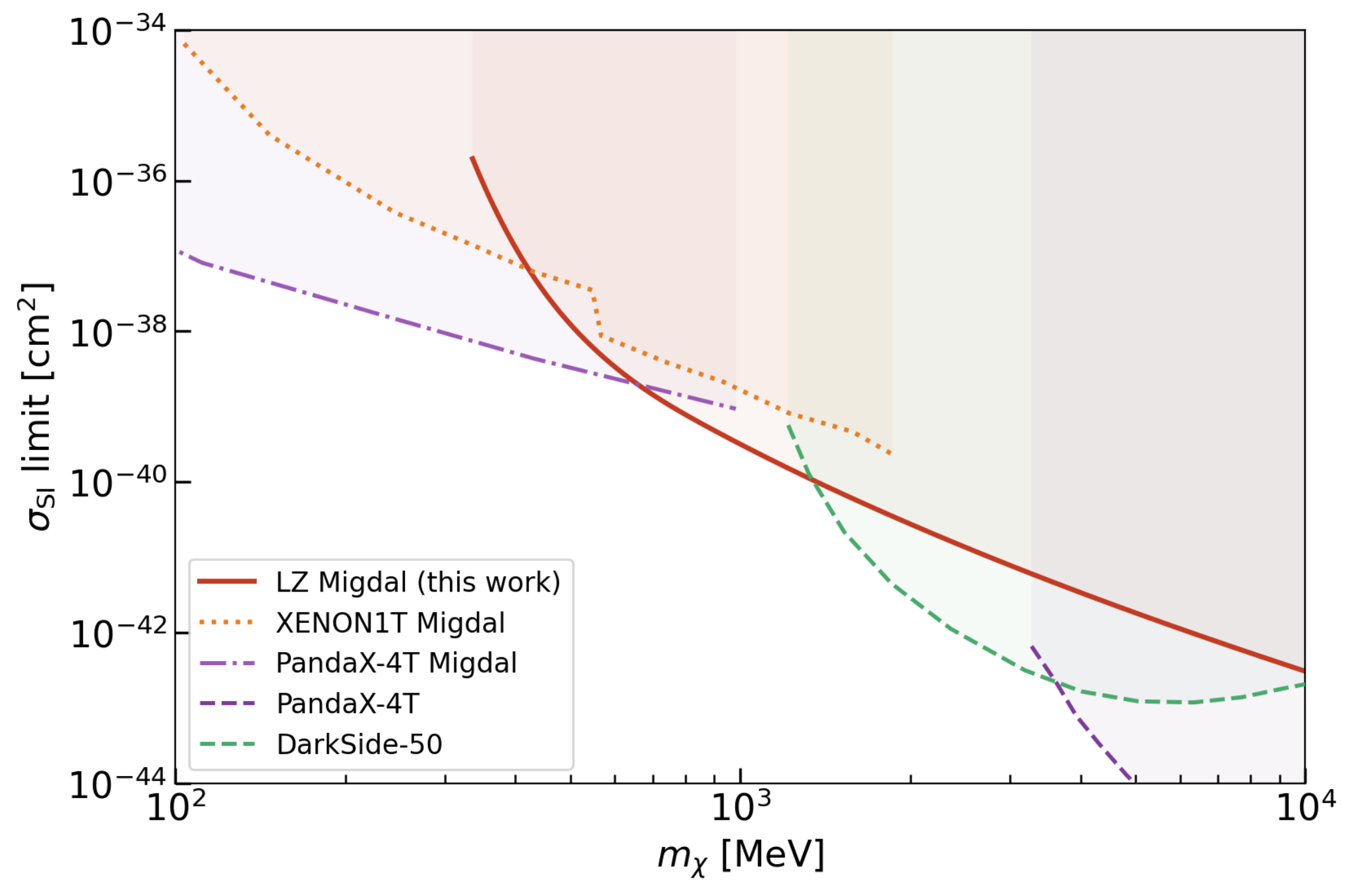}
  \caption{Spin-independent Migdal limit from the LZ keV templates compared with the PandaX, XENON and DarkSide limits.}
  \label{fig:lz-limit}
\end{figure}

The LZ Migdal limit in the $\sigma_{\mathrm{SI}}$--$m_{\chi}$ plane is
shown in Fig.~\ref{fig:lz-limit}, together with the Migdal limits from
PandaX-4T~\cite{Qiao:2023pbw} and XENON1T~\cite{XENON1T2019},
and the conventional nuclear-recoil limits from
PandaX-4T~\cite{PandaX4T2023} and DarkSide-50~\cite{DarkSide-50:2022qzh}.
The liquid-xenon target of LZ shares a similar atomic-structure context with XENON1T and PandaX-4T, whereas the liquid-argon target of DarkSide-50 provides different nuclear masses and electronic shell conditions~\cite{Grilli2020}.
The differences in onset masses and slopes of these limits reflect the combined differences in target material, threshold, exposure, detector response, and background model.

For DM masses below $\sim350\,\mathrm{MeV}$, no particle in the standard halo deposits enough energy to produce a Migdal event within the LZ analysis window. 
Consequently, the window-integrated signal templates vanish identically, and no finite cross-section upper limit can be derived.
Just above this threshold, the limit remains weak because only the extreme high-velocity tail of the halo velocity distribution contributes. 
As the DM mass increases, the accessible recoil-energy and velocity phase space opens up, and the limit improves rapidly. 
Around $ 1\,\mathrm{GeV}$, the LZ constraint surpasses the other limits considered and becomes the strongest constraint in this region.
For larger $m_\chi$, conventional nuclear-recoil experiments provide the strongest constraints.

The dominant systematic uncertainties arise from the Migdal ionization probabilities, the liquid-xenon electronic-recoil response, the energy calibration, the detection efficiency, and the background spectral shape~\cite{LZ2023ER}. 
The ionization probabilities require a many-electron atomic-structure calculation, while the visible energy also depends on the de-excitation and charge--photon production models~\cite{Cox2023}.
Background normalization and spectral-shape uncertainties are incorporated into the likelihood through nuisance parameters. 
Variations in the DM velocity-distribution parameters can shift the low-mass limits by $\mathcal{O}(10\text{--}100\%)$.

\section{Constraints on light mediator models}\label{sec:mediator} 

The limits derived in Sec.~\ref{sec:migdal} are obtained in the contact-interaction limit, corresponding to $m_{\mathrm{med}}\gg q$ where the mediator can be integrated out.
For a concrete DM model with a mediator, the scattering cross section is momentum dependent, and the limits must be reinterpreted in the model's parameter space.
We therefore convert each contact-interaction limit into a bound on the coupling combination $\alpha_{\mathrm{med}} = g_N g_\chi/(4\pi)$ using the following relation (see Appendix~\ref{app:conversion} for a detailed derivation)
\begin{align}
   \alpha_{med}(m_{med}) &= \frac{m_{med}^2}{4\mu_{\chi p}}
  \sqrt{\frac{\sigma_{SI}\,}{\pi}}\,
  \frac{1}{\sqrt{\langle f\rangle}} ,
  \label{alpha}
\end{align}
where $\sigma_{SI}$ is the constant contact cross section that reproduces the same event count, $\langle f\rangle = {\sum_i R(E_{R,i})\, f(E_{R,i})}/{\sum_i R(E_{R,i})} $ is the event-weighted fraction of the contact rate surviving the propagator, and $R(E_R)$ is the Migdal signal spectrum in the analysis window.

\begin{figure}[htbp]
  \centering
  \includegraphics[width=0.45\textwidth]{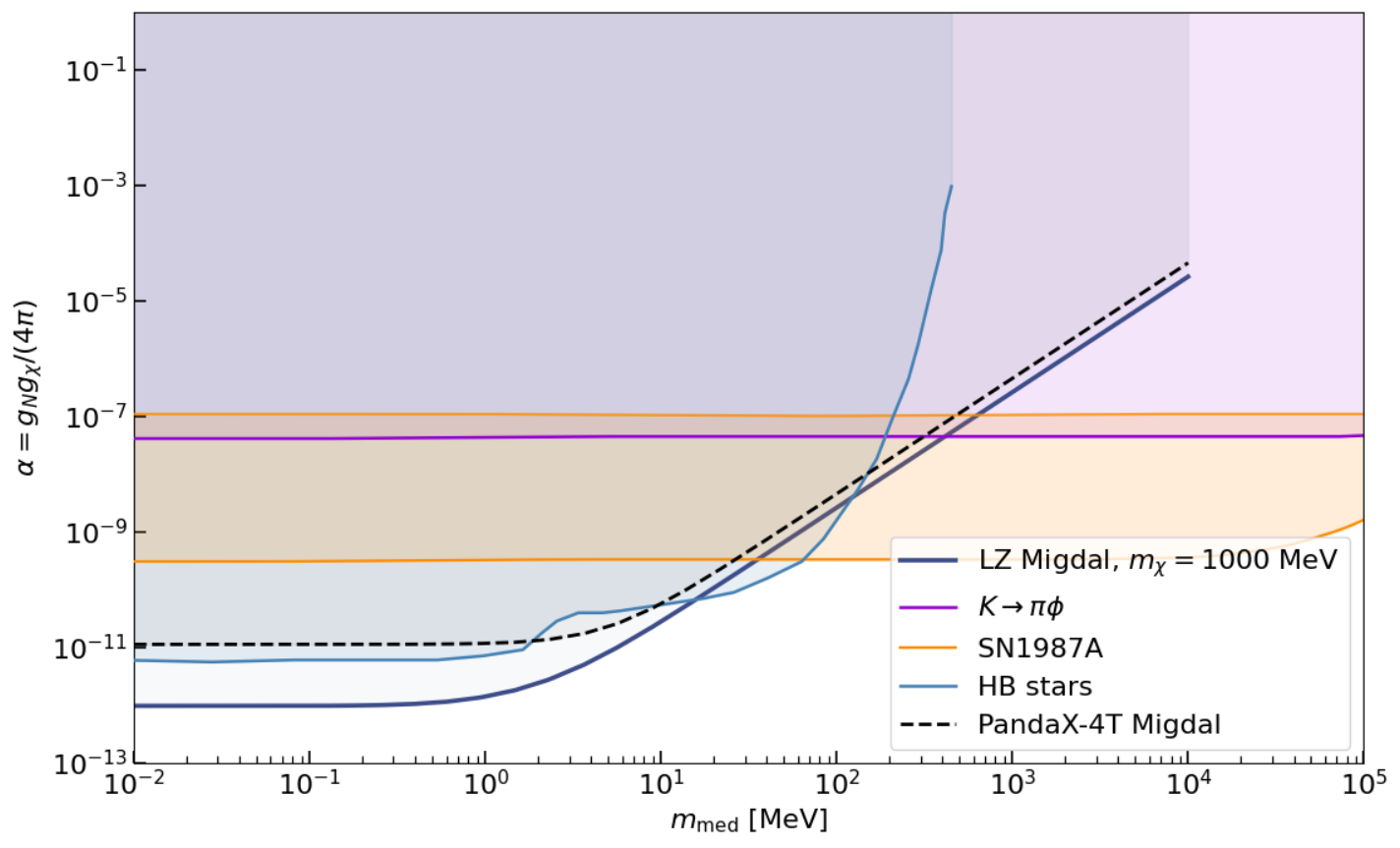}
  \includegraphics[width=0.45\textwidth]{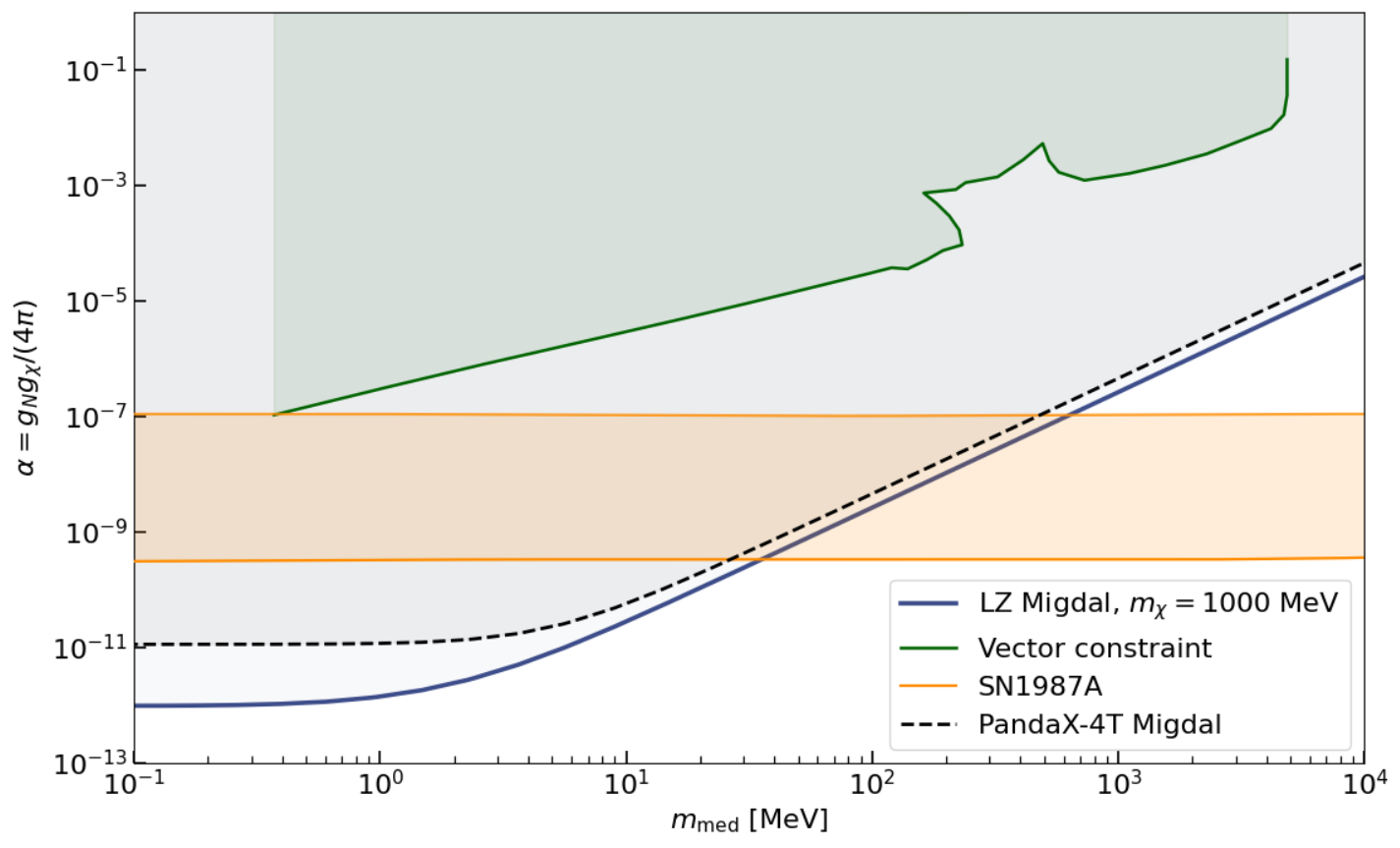}
  \caption{ LZ Migdal $\alpha_{med}$ upper limit ($m_{\chi}=1\,\mathrm{GeV}$) for scalar (\textit{Top}) and vector mediators (\textit{Bottom}).}  
  \label{fig:alpha-mediators}
\end{figure}

For the representative mass $m_{\chi}=1\,\mathrm{GeV}$, the two panels of Fig.~\ref{fig:alpha-mediators} show the LZ Migdal limit on $\alpha_{\mathrm{med}}$ for scalar (top) and vector (bottom) mediators, together with the other constraints in the same plane. 
We emphasize that this comparison requires consistent assumptions about the mediator spin, its couplings, and its production in each environment, with each bound mapped onto the same universal, isospin-independent nucleon coupling~\cite{Bell2024}. 
The derivation of the other constraints shown in Fig.~\ref{fig:alpha-mediators} relies on additional model details, including an invisibly decaying scalar for the $K\to\pi\phi$ limit~\cite{E949}, photon mixing for the HB bound~\cite{HardyLasenby}, the production channel and trapping for SN1987A~\cite{Chang2017,RrapajReddy2016}, and the vector-mediator constraint of Ref.~\cite{Dror2017}.



Since the leading non-relativistic operator is common to the scalar and vector mediator models, the LZ limit is identical in the two cases, while the two panels differ only in the external constraints that apply.
At low mediator masses, the LZ limit reaches about $10^{-12}$, which is much stronger than those from PandaX-4T and other experiments. 
With increasing $m_{\mathrm{med}}$, the two limits gradually converge toward each other, and their ratio approaches a constant.
For the scalar mediator model, the LZ Migdal-effect limit is the most stringent for $m_{\mathrm{med}}\lesssim10\,\mathrm{MeV}$. 
It weakens with increasing mediator mass and eventually becomes weaker than the HB stellar-cooling bound.
In the vector-mediator case, the other constraints are considerably weaker, and LZ provides the most stringent bound for $m_{\mathrm{med}}\lesssim10\,\mathrm{MeV}$.

\section{Conclusions}\label{sec:conclusion}
The Migdal effect provides a visible electronic signal that complements the conventional nuclear-recoil channel, extending the reach of direct detection to low-mass DM. 
In this work, we study the expected sensitivity of the LZ liquid-xenon detector to low-mass DM through this effect. 
Compared with other experiments such as DarkSide-50, PandaX-4T, and XENON1T, the LZ limit is the strongest for DM masses around $1\,\mathrm{GeV}$, which can be attributed to differences in experimental materials, thresholds, exposure, and analysis methods.
Furthermore, within the finite-mass mediator framework, we convert the cross-section limits into limits on the light-mediator DM model.  
Taking $m_{\chi}=1\,\mathrm{GeV}$ as representative, the LZ Migdal constraint on the coupling parameter at low mediator masses is stronger than the external bounds by one to two orders of magnitude.
These results show that the Migdal effect is an effective method for searching for low-mass DM particles and provide strong limits on the DM model parameter space.

\section*{Acknowledgments}
This work is supported by the National Key R\&D Program of China (Grant No. 2022YFF0503304), the National Natural Science Foundation of China (Grant Nos. 12373002, 12220101003).

\newpage
\appendix

\section{Derivation of the cross-section--coupling conversion}
\label{app:conversion}

In this appendix, we derive the conversion between the contact cross-section limit and the mediator coupling used in Eq.~(\ref{alpha}),  and clarify the origins of the propagator factor and the event-weighted average. Natural units are used throughout.


For DM scattering on a nucleus via a scalar or vector mediator of mass
$m_{\mathrm{med}}$, the tree-level amplitude includes the mediator
propagator, so that
$|\mathcal{M}|^2\propto(m_{\mathrm{med}}^2+q^2)^{-2}$.
The corresponding DM--nucleon cross section at momentum transfer $q$
is then
\begin{equation}\label{eq:app-sigmaq}
  \sigma(q)=\frac{(g_Ng_\chi)^2\mu_{\chi p}^2}
  {\pi(m_{\mathrm{med}}^2+q^2)^2},
\end{equation}
where $\mu_{\chi p}=m_\chi m_p/(m_\chi+m_p)$ is the DM--nucleon reduced mass, and the nucleus-level cross section is obtained by multiplying $\sigma(q)$ by the coherent factor, $\sigma_N(q)=A^2(\mu_{\chi N}/\mu_{\chi p})^2\sigma(q)$.
In the contact limit, $q^2\ll m_{\mathrm{med}}^2$,
\begin{equation}\label{eq:app-sigmac}
  \sigma_c\equiv\sigma(q\to0)=\frac{(g_Ng_\chi)^2\mu_{\chi p}^2}{\pi m_{\mathrm{med}}^4},
\end{equation}
while at finite $q$ the cross section is suppressed by the propagator factor
\begin{equation}\label{eq:app-f}
  \begin{split}
    f(q)&\equiv\frac{\sigma(q)}{\sigma_c}
    =\left(\frac{m_{\mathrm{med}}^2}{m_{\mathrm{med}}^2+q^2}\right)^2\in(0,1],
  \end{split}
\end{equation}
so that $\sigma(q)=\sigma_c\,f(q)$, with $q^2=2m_NE_R$ for a nuclear recoil of energy $E_R$. The coherent factor cancels in the ratio. 

Defining $\alpha_{\mathrm{med}}\equiv g_Ng_\chi/(4\pi)$,  Eq.~\eqref{eq:app-sigmac} can then be inverted to give the contact-limit conversion
\begin{equation}\label{eq:app-alphac}
  \begin{split}
    \sigma_c=\frac{16\pi\alpha_{\mathrm{med}}^2\mu_{\chi p}^2}{m_{\mathrm{med}}^4}.\\
  \end{split}
\end{equation}
Solving Eq.~\eqref{eq:app-alphac} for $\alpha_{\mathrm{med}}$ gives the coupling as a function of $\sigma_c$,
\begin{equation}\label{eq:app-alphac2}
  \begin{split}
    \alpha_{\mathrm{med}}=\frac{m_{\mathrm{med}}^2}{4\mu_{\chi p}}\sqrt{\frac{\sigma_c}{\pi}} .
  \end{split}
\end{equation}

The cross-section limit $\sigma_{\mathrm{SI}}^{90}$ denotes the $90\%$ CL upper bound on a \emph{constant} contact cross section obtained in Sec.~\ref{sec:method}.
Let $R(E_R)$ denote the differential signal rate in the analysis window. For a constant contact cross section $\sigma$, the expected number of events is therefore $N_c(\sigma)\propto\sigma\sum_i R(E_{R,i})$. For a finite-mass mediator with contact value $\sigma_c$, the rate is instead proportional to $\sigma(q)=\sigma_c f(q)$, giving
\begin{equation}\label{eq:app-N}
  \begin{split}
    N(\sigma_c)&\propto\sum_i R(E_{R,i})\,\sigma(q_i)
    =\sigma_c\sum_i R(E_{R,i})\,f(q_i)\\
    &=\sigma_c\,\langle f\rangle\sum_i R(E_{R,i})
    =\langle f\rangle\,N_c(\sigma_c),
  \end{split}
\end{equation}
where $q_i\equiv q(E_{R,i})$ and $q_i^2=2m_NE_{R,i}$. The event-weighted average is therefore
\begin{equation}\label{eq:app-favg}
  \langle f\rangle
  =\frac{\sum_i R(E_{R,i})\,f(q_i)}{\sum_i R(E_{R,i})}.
\end{equation}
Requiring the finite-mass model to reproduce the event count predicted by the contact model at its $90\%$ CL limit, $N(\sigma_c)=N_c(\sigma_{\mathrm{SI}}^{90})$, then determines the contact cross section of the finite-mass model as
\begin{equation}\label{eq:app-sigmac2}
  \sigma_c=\frac{\sigma_{\mathrm{SI}}^{90}}{\langle f\rangle}.
\end{equation}
Substituting Eq.~\eqref{eq:app-sigmac2} into Eq.~\eqref{eq:app-alphac2} gives the conversion used in Eq.~(\ref{alpha})
\begin{align}
   \alpha_{med}(m_{med}) &= \frac{m_{med}^2}{4\mu_{\chi p}}
  \sqrt{\frac{\sigma_{SI}\,}{\pi}}\,
  \frac{1}{\sqrt{\langle f\rangle}}.
\end{align}





\begin{thebibliography}{99}
\bibitem{Rubin:1980zd}
V.~C.~Rubin, N.~Thonnard and W.~K.~Ford, Jr.,
Astrophys. J. \textbf{238} (1980), 471

\bibitem{Zwicky1937}
F.~Zwicky,
\emph{Astrophys. J.} \textbf{86}, 217 (1937),
\href{https://doi.org/10.1086/143864}{doi:10.1086/143864}.

\bibitem{Planck2018}
N.~Aghanim \emph{et al.} (Planck Collaboration),
\emph{Astron. Astrophys.} \textbf{641}, A6 (2020),
\href{https://arxiv.org/abs/1807.06209}{arXiv:1807.06209}.


\bibitem{Clowe:2006eq}
D.~Clowe, M.~Bradac, A.~H.~Gonzalez, M.~Markevitch, S.~W.~Randall, C.~Jones and D.~Zaritsky,
Astrophys. J. Lett. \textbf{648} (2006), L109-L113
[arXiv:astro-ph/0608407 [astro-ph]].



\bibitem{GoodmanWitten1985}
M.~W.~Goodman and E.~Witten,
\emph{Phys. Rev. D} \textbf{31}, 3059 (1985).


\bibitem{UndagoitiaRauch2016}
T.~Marrod\'an Undagoitia and L.~Rauch,
\emph{J. Phys. G} \textbf{43}, 013001 (2016),
\href{https://arxiv.org/abs/1509.08767}{arXiv:1509.08767}.

\bibitem{BertoneHooperSilk2005}
G.~Bertone, D.~Hooper, and J.~Silk,
\emph{Phys. Rep.} \textbf{405}, 279 (2005),
\href{https://arxiv.org/abs/hep-ph/0404175}{arXiv:hep-ph/0404175}.

\bibitem{Billard2022}
J.~Billard \emph{et al.},
\emph{Rep. Prog. Phys.} \textbf{85}, 056201 (2022),
\href{https://arxiv.org/abs/2104.07634}{arXiv:2104.07634}.

\bibitem{LZ2023}
J.~Aalbers \emph{et al.} (LZ Collaboration),
\emph{Phys. Rev. Lett.} \textbf{131}, 041002 (2023),
\href{https://arxiv.org/abs/2207.03764}{arXiv:2207.03764}.

\bibitem{LZ2025}
J.~Aalbers \emph{et al.} (LZ Collaboration),
\emph{Phys. Rev. Lett.} \textbf{135}, 011802 (2025),
\href{https://arxiv.org/abs/2410.17036}{arXiv:2410.17036}.

\bibitem{XENONnT2023}
E.~Aprile \emph{et al.} (XENON Collaboration),
\emph{Phys. Rev. Lett.} \textbf{131}, 041003 (2023),
\href{https://arxiv.org/abs/2303.14729}{arXiv:2303.14729}.

\bibitem{PandaX4T2021}
Y.~Meng \emph{et al.} (PandaX-4T Collaboration),
\emph{Phys. Rev. Lett.} \textbf{127}, 261802 (2021),
\href{https://arxiv.org/abs/2107.13438}{arXiv:2107.13438}.

\bibitem{OHare2021}
C.~A.~J.~O'Hare,
\emph{Phys. Rev. Lett.} \textbf{127}, 251802 (2021),
\href{https://arxiv.org/abs/2109.03116}{arXiv:2109.03116}.

\bibitem{BringmannPospelov2019}
T.~Bringmann and M.~Pospelov,
\emph{Phys. Rev. Lett.} \textbf{122}, 171801 (2019),
\href{https://arxiv.org/abs/1810.10543}{arXiv:1810.10543}.

\bibitem{Yu:2025bwu}
Y.~Yu, G.~S.~Wang, B.~Zhang, T.~P.~Tang, B.~Y.~Su and L.~Feng,
Eur. Phys. J. C \textbf{86} (2026) no.10, 1164
[arXiv:2512.23282 [astro-ph.HE]].

\bibitem{AnPospelovPradlerRitz2018}
H.~An, M.~Pospelov, J.~Pradler, and A.~Ritz,
\emph{Phys. Rev. Lett.} \textbf{120}, 141801 (2018),
\href{https://arxiv.org/abs/1708.03642}{arXiv:1708.03642}.

\bibitem{Wang:2023sxr}
G.~S.~Wang, Z.~F.~Chen, L.~Zu, H.~Gong, L.~Feng and Y.~Z.~Fan,
JCAP \textbf{05} (2024), 129
[arXiv:2303.14117 [astro-ph.HE]].


\bibitem{Wang:2026khh}
G.~S.~Wang, B.~Y.~Su, Y.~Yu, B.~Zhang and L.~Feng,
\emph{Phys. Rev. D} \textbf{114}, 043022 (2026),
\href{https://arxiv.org/abs/2602.08731}{arXiv:2602.08731 [astro-ph.CO]}.


\bibitem{Madhavacheril2014}
M.~S.~Madhavacheril, N.~Sehgal, and T.~R.~Slatyer,
\emph{Phys. Rev. D} \textbf{89}, 103508 (2014),
\href{https://arxiv.org/abs/1310.3815}{arXiv:1310.3815}.


\bibitem{Damico2018}
G.~D'Amico, A.~Panci, and A.~Strumia,
\emph{Phys. Rev. Lett.} \textbf{121}, 011103 (2018),
\href{https://arxiv.org/abs/1803.03629}{arXiv:1803.03629}.



\bibitem{Migdal1941}
A.~B.~Migdal,
\emph{J. Phys. (USSR)} \textbf{4}, 449 (1941).

\bibitem{Ibe2018}
M.~Ibe, W.~Nakano, Y.~Shoji, and K.~Suzuki,
\emph{JHEP} \textbf{03}, 194 (2018),
\href{https://arxiv.org/abs/1707.07258}{arXiv:1707.07258}.

\bibitem{Bell2022}
N.~F.~Bell, J.~B.~Dent, R.~F.~Lang, J.~L.~Newstead, and A.~C.~Ritter,
\emph{Phys. Rev. D} \textbf{105}, 096015 (2022),
\href{https://arxiv.org/abs/2112.08514}{arXiv:2112.08514}.

\bibitem{Xu:2023wev}
J.~Xu, D.~Adams, B.~Lenardo, T.~Pershing, R.~Mannino, E.~Bernard, J.~Kingston, E.~Mizrachi, J.~Lin and R.~Essig, \textit{et al.}
Phys. Rev. D \textbf{109} (2024) no.5, L051101
[arXiv:2307.12952 [hep-ex]].

\bibitem{XENON1T2019}
E.~Aprile \emph{et al.} (XENON Collaboration),
\emph{Phys. Rev. Lett.} \textbf{123}, 241803 (2019),
\href{https://arxiv.org/abs/1907.12771}{arXiv:1907.12771}.


\bibitem{DarkSide:2022dhx}
P.~Agnes \textit{et al.} [DarkSide],
Phys. Rev. Lett. \textbf{130} (2023) no.10, 10
[arXiv:2207.11967 [hep-ex]].



\bibitem{Qiao:2023pbw}
M.~Qiao, C.~Xia and Y.~F.~Zhou,
JCAP \textbf{11} (2023), 079
[arXiv:2307.12820 [hep-ph]].

\bibitem{Scopel:2026aqk}
S.~Scopel and G.~Tomar,
[arXiv:2608.18609 [hep-ph]].

\bibitem{Yi2026}
D.~Yi \emph{et al.},
\emph{Nature} \textbf{649}, 580 (2026),
\href{https://doi.org/10.1038/s41586-025-09918-8}{doi:10.1038/s41586-025-09918-8}.


\bibitem{Dolan2018}
M.~J.~Dolan, F.~Kahlhoefer, and C.~McCabe,
\emph{Phys. Rev. Lett.} \textbf{121}, 101801 (2018),
\href{https://arxiv.org/abs/1711.09906}{arXiv:1711.09906}.

\bibitem{LewinSmith}
J.~D.~Lewin and P.~F.~Smith,
\emph{Astropart. Phys.} \textbf{6}, 87 (1996).

\bibitem{Helm1956}
R.~H.~Helm,
\emph{Phys. Rev.} \textbf{104}, 1466 (1956).

\bibitem{Cox2023}
P.~Cox, M.~J.~Dolan, C.~McCabe, and H.~M.~Quiney,
\emph{Phys. Rev. D} \textbf{107}, 035032 (2023),
\href{https://arxiv.org/abs/2208.12222}{arXiv:2208.12222}.

\bibitem{PandaX4T2023}
S.~Li \emph{et al.} (PandaX-4T Collaboration),
\emph{Phys. Rev. Lett.} \textbf{130}, 261001 (2023),
\href{https://arxiv.org/abs/2212.10067}{arXiv:2212.10067}.

\bibitem{DarkSide-50:2022qzh}
P.~Agnes \textit{et al.} [DarkSide-50],
Phys. Rev. D \textbf{107} (2023) no.6, 6
[arXiv:2207.11966 [hep-ex]].

\bibitem{Grilli2020}
G.~Grilli~di~Cortona, A.~Messina, and S.~Piacentini,
\emph{JHEP} \textbf{11}, 034 (2020),
\href{https://arxiv.org/abs/2006.02453}{arXiv:2006.02453}.

\bibitem{LZ2023ER}
J.~Aalbers \emph{et al.} (LZ Collaboration),
\emph{Phys. Rev. D} \textbf{108}, 072006 (2023),
\href{https://arxiv.org/abs/2307.15753}{arXiv:2307.15753}.

\bibitem{Bell2024}
N.~F.~Bell, J.~L.~Newstead, and I.~Shaukat-Ali,
\emph{Phys. Rev. D} \textbf{109}, 063034 (2024),
\href{https://arxiv.org/abs/2309.11003}{arXiv:2309.11003}.

\bibitem{E949}
A.~V.~Artamonov \emph{et al.} (E949 Collaboration),
\emph{Phys. Rev. D} \textbf{79}, 092004 (2009).

\bibitem{HardyLasenby}
E.~Hardy and R.~Lasenby,
\emph{JHEP} \textbf{02}, 033 (2017),
\href{https://arxiv.org/abs/1611.05852}{arXiv:1611.05852}.

\bibitem{Chang2017}
J.~H.~Chang, R.~Essig, and S.~D.~McDermott,
\emph{JHEP} \textbf{01}, 107 (2017),
\href{https://arxiv.org/abs/1611.03864}{arXiv:1611.03864}.

\bibitem{RrapajReddy2016}
E.~Rrapaj and S.~Reddy,
\emph{Phys. Rev. C} \textbf{94}, 045805 (2016),
\href{https://arxiv.org/abs/1511.09136}{arXiv:1511.09136}.

\bibitem{Dror2017}
J.~A.~Dror, R.~Lasenby, and M.~Pospelov,
\emph{Phys. Rev. Lett.} \textbf{119}, 141803 (2017),
\href{https://arxiv.org/abs/1705.06726}{arXiv:1705.06726}.

\bibitem{LeeWeinberg1977}
B.~W.~Lee and S.~Weinberg,
\emph{Phys. Rev. Lett.} \textbf{39}, 165 (1977).

\bibitem{PospelovRitzVoloshin2008}
M.~Pospelov, A.~Ritz, and M.~B.~Voloshin,
\emph{Phys. Lett. B} \textbf{662}, 53 (2008),
\href{https://arxiv.org/abs/0711.4866}{arXiv:0711.4866}.

\bibitem{PandaX4TMed}
D.~Huang \emph{et al.} (PandaX-4T Collaboration),
\emph{Phys. Rev. Lett.} \textbf{131}, 191002 (2023),
\href{https://arxiv.org/abs/2308.01540}{arXiv:2308.01540}.

\bibitem{BoehmFayet2004}
C.~Boehm and P.~Fayet,
\emph{Nucl. Phys. B} \textbf{683}, 219 (2004),
\href{https://arxiv.org/abs/hep-ph/0305261}{arXiv:hep-ph/0305261}.

\bibitem{Essig2013}
R.~Essig \emph{et al.},
\href{https://arxiv.org/abs/1311.0029}{arXiv:1311.0029} (2013).


\end{thebibliography}
\end{document}